\documentclass[10pt,conference]{IEEEtran}
\usepackage[utf8]{inputenc}
\usepackage{times}
\usepackage{verbatim}
\usepackage[noadjust]{cite}
\usepackage{graphicx}
\usepackage{setspace}
\usepackage{color}
\usepackage{authblk}
\usepackage{commath}
\usepackage{tabulary}
\graphicspath{ {Experimental_Results/} }
\usepackage{amsmath}
\usepackage{algorithm}
\usepackage{algorithmic}
\usepackage[figurename=Figure,labelsep=period,tablename=Table]{caption}
\usepackage{xr}
\usepackage{textcomp}
\usepackage[normalem]{ulem}
\usepackage{enumitem}
\setlist{leftmargin=3mm}

\providecommand{\keywords}[1]{\textbf{\textit{Keywords---}}#1}

\author {
	{{Pankaj Saha},{Angel Beltre}, and {Madhusudhan Govindaraju}}
	\vspace{1.6mm}\\
	\fontsize{10}{10}\selectfont\itshape
    {Cloud and Big Data Laboratory},
	{State University of New York (SUNY) at Binghamton}\\
	\fontsize{9}{9}\selectfont\ttfamily\upshape
	{
		{{$\lbrace$psaha4, abeltre1, mgovinda$\rbrace$@binghamton.edu }}
	}
}

\begin{document}

\def\sharedaffiliation{%
\end{tabular}
\begin{tabular}{c}}

\title{Exploring the Fairness and Resource Distribution in an Apache Mesos Environment}

\renewcommand{\thetable}{\arabic{table}}
\date{}
\maketitle
\thispagestyle{empty}
\pagestyle{empty}

\begin{abstract}

Apache Mesos, a cluster-wide resource manager, is widely deployed in massive scale at several Clouds and Data Centers. Mesos aims to provide high cluster utilization via fine grained resource co-scheduling and resource fairness among multiple users through Dominant Resource Fairness (DRF) based allocation. DRF takes into account different resource types (CPU, Memory, Disk I/O) requested by each application and determines the share of each cluster resource that could be allocated to the applications. 

Mesos has adopted a two-level scheduling policy: (1) DRF to allocate resources to competing frameworks and (2) task level scheduling by each framework for the resources allocated during the previous step.
We have conducted experiments in a local Mesos cluster when used with frameworks such as Apache Aurora, Marathon, and our own framework Scylla, to study resource fairness and cluster utilization. 

Experimental results show how informed decision regarding second level scheduling policy of frameworks and attributes like offer holding period, offer refusal cycle and task arrival rate can reduce unfair resource distribution. Bin-Packing scheduling policy on Scylla with Marathon can reduce unfair allocation from 38\% to 3\%. By reducing unused free resources in offers we bring down the unfairness from to 90\% to 28\%. We also show the effect of task arrival rate to reduce the unfairness from 23\% to 7\%. 
\footnote{
This work was supported in part by NSF grant OAC-1740263.}

\end{abstract}

\section{Introduction}
Widely known fair resource sharing policies such as max-min fairness and the generalized variation, weighted max-min fairness, are designed  to provide a {\tt fair share guarantee}~\cite{wang2014dominant}. However, these only work satisfactorily when a single resource type, such as CPU or memory, is taken into account. In data centers and clouds, where applications could be co-scheduled on the same physical nodes, resource fairness needs to extend to multiple resource types such as memory, disk I/O, and network bandwidth. The Dominant Resource Fairness (DRF)~\cite{ghodsi2011dominant} algorithm was introduced to address this requirement for resource management of large-clusters and cloud environments. 

The key tenant of DRF is that it takes into account different resource types (CPU, Memory, Disk I/O) requested by each application and determines the share of each cluster resource that could be allocated to the applications. DRF has been adopted by Apache Mesos \cite{Hindman2011Mesos:Center}, a widely used cluster-wide operating system that can efficiently manage very large clusters. 
Mesos is estimated to seamlessly scale to more than 10K nodes in some commercial settings~\cite{MesosNews}. 

Mesos pools all the resources in a cluster and allows fine-grained resource sharing by allowing and enforcing multiple applications (called Mesos frameworks) to co-schedule their tasks on VMs/nodes. Mesos employs DRF to allocate resources to frameworks, and then the frameworks use scheduling algorithms to schedule tasks within the allocated resources. The frameworks that are widely deployed to work in concert with Apache Mesos are Apache Aurora \cite{ApacheAurora} for long-running services,  Mesosphere Marathon~\cite{Marathon:DC/OS} for container orchestration, and Chronos~\cite{Chronos:Mesos} for cron jobs. In previous work, we developed a Mesos framework, Scylla~\cite{Saha2017Scylla:Jobs}, for MPI based HPC jobs. 

 
 
While DRF is well intentioned, there are several use cases where the Mesos framework's internal scheduling policy, and attribute settings that govern interaction with DRF, prevent it from meeting the desired fairness objectives. In this paper, we have identified a few key attributes in a framework that affect access to a fair share of resources, when used in an Apache Mesos cluster. These include interaction with the Mesos resource offer cycles, offer holding period, task arrival rate, and task duration. 
We provide suggestions for how cluster administrators can control these attributes to ensure fair distribution of resources across all users of a cluster.

We assume that cluster managers and framework developers will use {\tt off-the-shelf} Apache Mesos and its DRF modules, as it has the advantage of receiving support from service providers and the developer community, and can leverage seamless upgrades to the core Mesos tools. Modifying the DRF implementation within Mesos requires cluster managers and framework developers to maintain custom versions and manually keep track and incorporate the patches to Mesos and DRF modules.


The key contributions of this paper are the following: 

\begin{itemize} 
\item We have identified the differences between the well-known DRF algorithm and its variation that is available with the widely used Apache Mesos distribution.  

\item We have studied and analyzed the behavior of an Apache Mesos based cluster and determined the key attributes that control the resource distribution among multiple users.

\item We have developed recommendations for cluster administrators for configuring key attributes to significantly improve resource distribution among all the frameworks, for varying workloads.


\end{itemize}

\section{Motivation}

\begin{figure}[h!]
 \vspace{-1.5em}
  \includegraphics[width=0.5\textwidth,height=4.5cm]  {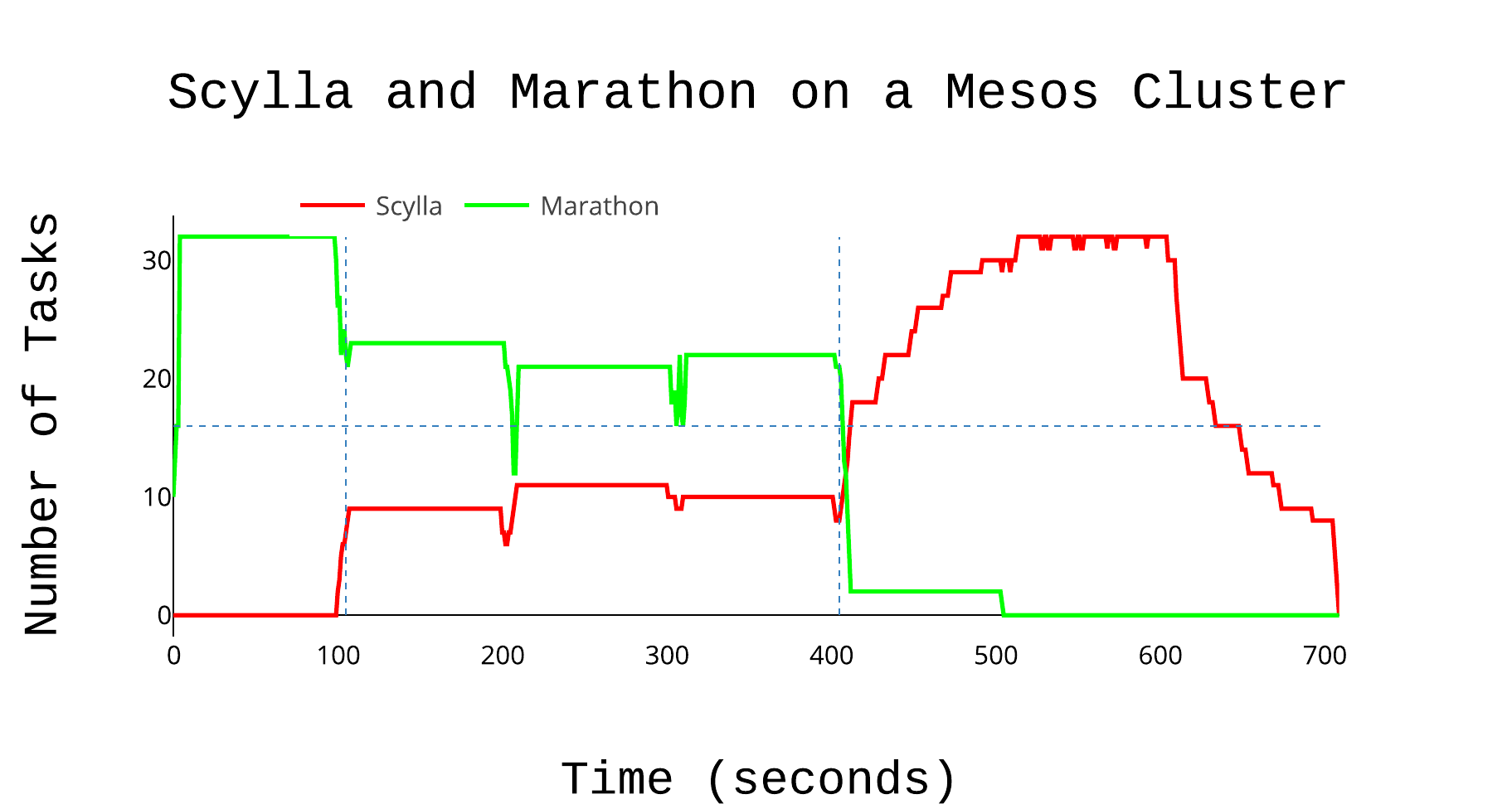}
  \caption{{\it Unfair Distribution: Scylla and Marathon competing for resources in a Mesos cluster. Marathon gets significantly more resources and is thus able to launch several more tasks than Scylla. Scylla's tasks face long wait times due to unfair distribution.}
   }
  \label{ScyllaVsMarathon}
   \vspace{-1.5em}
\end{figure}

\begin{figure}[h!]
\vspace{-0.1em}
  \includegraphics[width=0.5\textwidth,height=4.5cm]
  {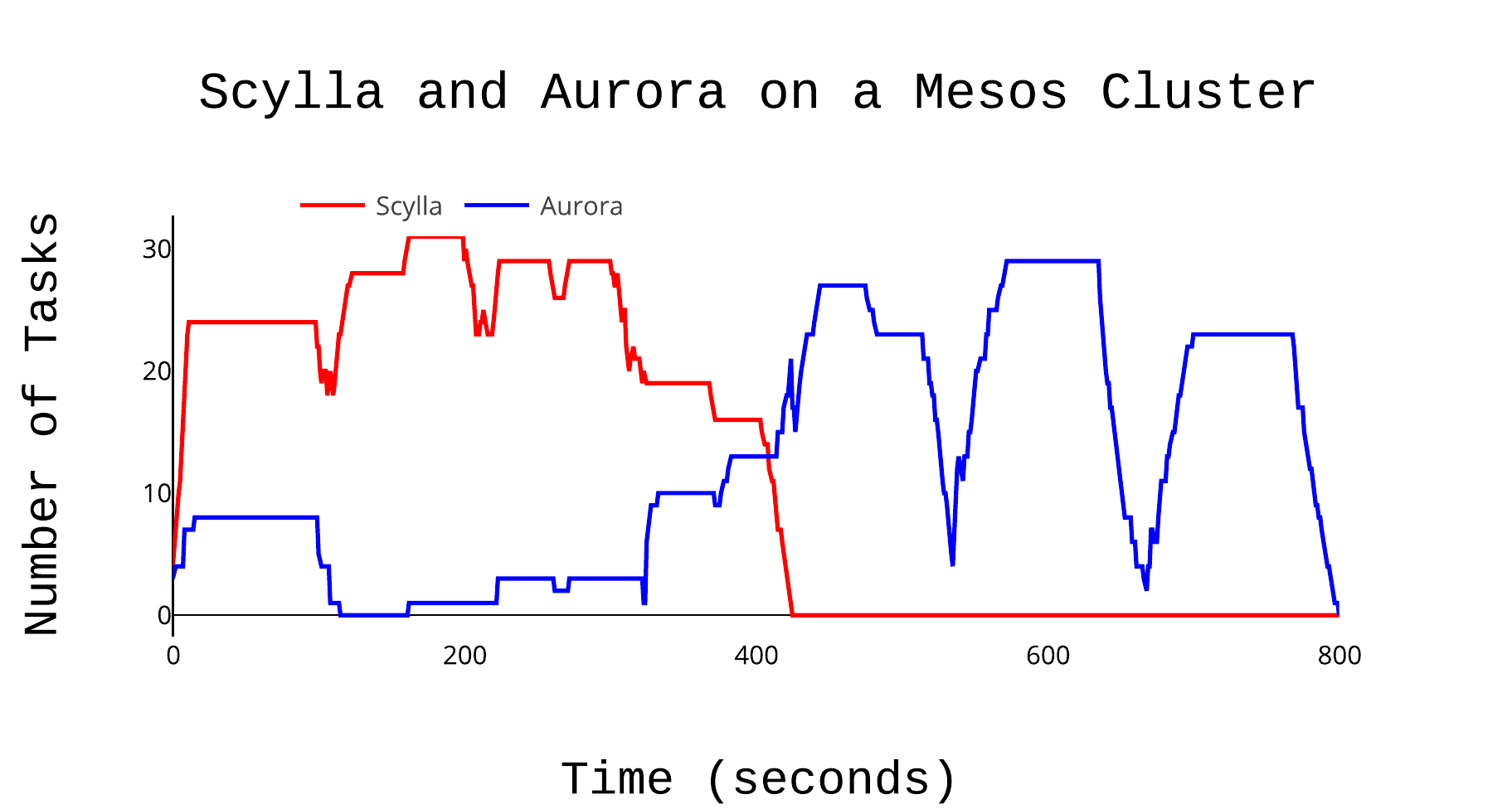}
  \caption{{\it Starvation: Scylla and Aurora competing for resources in a Mesos cluster. Scylla gets significantly more resources in the beginning. Aurora is able to launch its tasks only after Scylla is done launching all its tasks. Aurora's tasks are starving for resources when launched along with Scylla.}
}
  \label{ScyllaVsAurora}
   \vspace{-0.7em}
\end{figure}

Figure \ref{ScyllaVsMarathon} and \ref{ScyllaVsAurora} show how the presence of a pair of frameworks can result in an unfair resource distribution, when the frameworks are launched without DRF and Mesos awareness. The frameworks have been given 100 tasks, each with  requirements of 1 CPU and 1 GB of RAM. The two frameworks are launched on a cluster with 32 CPUs and 64 GB RAM. 
In this experiment, as all tasks are identical in terms of resource demands, the number of tasks per second shows resource distribution across the frameworks. We have observed, over repeated runs, that with the default configuration of Marathon, Aurora, and Scylla, the resource distribution never converges to a fair share distribution. Either Aurora (in Figure \ref{ScyllaVsAurora}) or Scylla (in Figure \ref{ScyllaVsMarathon}) faces starvation, and is only able to launch tasks after Scylla does not have any pending tasks. Even though the Mesos Master tries to distribute the resource in a fair way, the individual framework’s configurations and scheduling policy affects the overall distribution. 

Quantifying Unfairness:
We measure the unfairness \text{$U_{A}$} to framework A by using the following formula.
\[ U_{A} = (\dfrac{Area_{i,j} \ by \ framework_{A}}{{Area_{i,j} \ by \ fair \ graph}})*100\]

\( Area_{i,j} \) is the area under the curve from point i to j

In Figure \ref{ScyllaVsMarathon}, the horizontal dotted line is the fair distribution threshold and the two vertical dotted lines are the start and end points where fairness is not achieved. The graph outside the starting and ending points are showing usage while either one framework has not started launching any tasks or done with launching all the tasks. In Figure \ref{ScyllaVsMarathon}, Scylla's fair share is reduced by 38\%, whereas in Figure~\ref{ScyllaVsAurora} Aurora's fair share is reduced by 67\%.

\section{Background}

\subsection{Apache Mesos Architecture}
Apache Mesos is composed of three primary components (1) Mesos Master, (2) Mesos Agent, and (3) Mesos Framework. The Mesos Master manages resources during resource negotiation between the Mesos Agents and Mesos frameworks. Mesos Agents are responsible for executing requested tasks with available resources. All the available free resources from a single Mesos agent is included in a resource offer. Mesos frameworks make their own scheduling decisions to map one or multiple tasks to the offers allocated to them.

\begin{figure}[h!]
 \vspace{-1.5em}
  \includegraphics[width=0.5\textwidth]
  {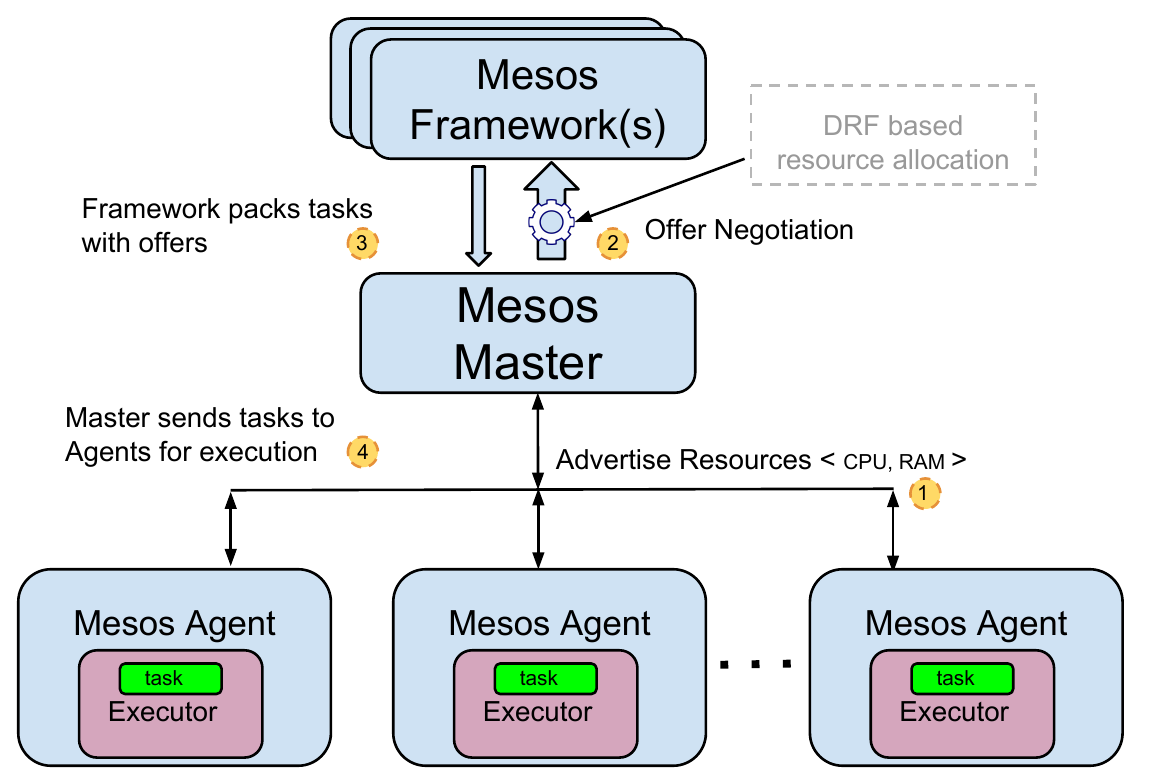}
  \caption{{\it Apache Mesos Architecture: This diagram shows the components of an Apache Mesos cluster and the steps involved in allocation of resources to frameworks }}
  \vspace{-0.5em}
  \label{mesos_diagram}
\end{figure}

In step 1, Mesos Agents periodically advertise to the Mesos Master regarding the available free resources (e.g. CPU, RAM, disk, I/O), which can be utilized for launching tasks. 
In step 2, Mesos Master’s allocation module starts offering the available resources to active frameworks. During each allocation cycle, Mesos Master sorts the frameworks based on the DRF algorithm and the framework with lowest dominant share is the first to receive the offer. After resource allocation, an individual framework can decide whether to accept or decline an offer. The declined offer goes back to the resource pool and will be offered to other frameworks in the next allocation cycle.
In step 3, if the offered resources satisfy a framework's resource demands, the framework creates a task list against offers. Based on the individual framework's policy, one or multiple tasks can be packed against a single resource offer.
In step 4, Mesos Master launches them on the chosen agents and if the required resources exceed the available resources in the offer, Mesos Master responds with an error. Otherwise, it sends a set of tasks to the individual agents corresponding to the offer.

\subsection{Mesos Framework}
A framework communicates with the Mesos Master for resource negotiation, and upon receiving the desired resources it executes its tasks on Mesos Agents. A Mesos Framework has two key modules: (1) Framework Scheduler and (2) Framework Executor. 

\subsubsection {Framework Scheduler} It is responsible for resource negotiations. The Mesos master decides how many offers will be given to each framework, but the framework's scheduling policy determines which offers will be picked among the available offers. Once offers are accepted, the framework's scheduler creates a map of tasks with the offers and then informs the Mesos Master to launch the tasks on the Mesos Agents associated with the offer. 
Apache Mesos implements a list of filters corresponding to each registered framework. After a framework rejects or partially uses an offer, it can prevent the same resources from being allocated to the framework by temporarily placing that agent in the filter list. The framework owner can configure the duration an agent should stay in the filter list. During that wait period, resources from the same agent are not offered. This wait time is known as {\tt Offer Refusal} and it is configured in units of seconds.

\subsubsection{Framework Executor} It is a process that resides on each Mesos Agent node and it is contacted whenever an agent node accepts tasks from frameworks. 

\subsection{Framework's Scheduling Policy} Apache Mesos' allocation module allocates resources to available frameworks based on the DRF fairness policy. The resources are listed in an {\tt offer} that has details of the agent node and the share of each resource type that has been allocated. Each framework scheduler can implement its custom scheduling policy.
Typically, a framework has a queue of tasks to schedule, and it uses a policy to decide how to pack tasks into offers. For example, {\tt Bin Packing} and {\tt First Fit} are two commonly used scheduling policies.

\section{Dominant Resource Fairness}
\subsection{How DRF works}
Apache Mesos provides two-level scheduling for resource allocation to frameworks. The Mesos Master’s allocation module decides the amount of resources that will be offered to each framework during every DRF offer cycle. The dominant resource of a framework is defined as the resource type that is used the most and is computed in terms of percentage of the overall availability of that resource. 
To calculate the dominant share \text{$S_{i}$} of user \text{$u_{i}$}, DRF uses the following formula:
\vspace{-0.5em}
\[ S_{i} = max_{j=1}^m  (\dfrac{u_{i,j}}{r _{j}})\]
 \vspace{-1.2em}
\[ m   = \text{available types of resources}\]
 \vspace{-1.8em}
\[ r_{j}  = \text{total available resources of type $j$} \]
 \vspace{-1.8em}
\[ u_{i,j}  = \text{amount of resource of type $j$, being use by user $u_{i}$} \]

Apache Mesos uses the DRF algorithm to decide the first level of resource distribution among the frameworks. In the second level of scheduling, a framework can choose which offer to accept and which one to decline. This second level of scheduling is pluggable and can be varied based on the requirements of the cluster. In each offer cycle, the Mesos Master tries to allocate resources through DRF to the framework that has the smallest dominant share. Then, it proceeds to offer to the second smallest share, and so on, until all the resources have been allocated. 
The DRF allocation module is not exercised in an environment where only one framework is in use. However, when multiple frameworks are competing, which is a common use case in very large clusters and data centers, Mesos uses DRF to attempt a fair allocation of resources based on each frameworks' current demands and usage. 

\subsection{DRF - Implementation in Mesos}

Once resources are allocated to a framework, it can choose which offers to accept and which ones to reject. The DRF implementation within Apache Mesos does not exactly follow the original algorithm. 


\subsubsection{Single Node vs Pool of Resources in Cluster}
The classical DRF algorithm~\cite{DominantBerkeley} is designed to allocate resources from a single node to competing users. However, Mesos pools resources from a heterogeneous set of nodes and presents a single view of the cluster-wide resources. After each offer has been allocated, which could be for resources spread across nodes, the Mesos Master recalculates the dominant share of all the users for allocating the next offer available in the cluster. The Master keeps allocating all the offers as long as resources are available in the cluster. Once all the available offers have been allocated and accepted, one cycle of allocation is considered to be completed.

\subsubsection{Resource Demands from Users}
The Mesos Master’s allocation module does not consider any demands from users. 
In the classical DRF algorithm, a demand (\(D_{i}\)) of user \((u_{i}\)) is considered as a vector \( D_{i}=<d_{i,1},d_{i,2},...,d_{i,m}> \) where 'm' is the number of resources in the cluster. A user receives resource offer as a multiple of the demand vector --  \( Offer_{i} \implies D_{i}*n=<d_{i,1}*n,d_{i,2}*n,...,d_{i,m}*n> \) where 'n' is the multiplication factor. This \( Offer_{i} \) is capable of launching 'n' tasks from user \(u_{i}\). 
Further, DRF assumes that all the tasks, received form a user, have identical resource demands.

However, unlike the classical DRF, Mesos Master allocates resources to users only based on the dominant share and offers all the available resources in an agent node. It allows users to  reject part of the offer, or the entire offer, based on the requirement.

\subsection{DRF-based Resource Allocation}
Apache Mesos Master offers resources to frameworks during each cycle of resource distribution and the framework with minimum dominant share is served first. After the framework uses a portion of the offered resources, the rejected and unused portion of the offer, if any, goes back to the resource pool and is eligible for allocation in the next cycle. After a resource is assigned to a framework, the Mesos Master resets the priority of the available frameworks based on their dominant share. A framework with the lowest dominant share holds the highest priority.


This resource allocation cycle runs against each unallocated offer in the cluster. For each unallocated offer, picked randomly from the list of offers, the allocation module finds the framework with the lowest dominant share. If the offer is in the filtered list of offers at that point of time, then the allocation module picks the next available framework in the DRF-sorted list of frameworks. Otherwise, it assign the resources to the framework.  


\begin{algorithm}
\caption{Resource Allocation Cycle in Mesos}\label{drf}
{\fontsize{10}{10}\selectfont
\begin{algorithmic}
 	\FOR{each agent in the randomSorted agents}
 		\FOR{each role in drfSorted roles}
 			\FOR{each framework in drfSorted frameworks}
 				\IF{framework already filtered these resources}
 					\STATE continue and skip the current framework
 				\ELSE{}
 					\STATE  allocate the resources to the framework
				\ENDIF
        	\ENDFOR
   		\ENDFOR
	\ENDFOR
\end{algorithmic}
}
\end{algorithm}
After each DRF cycle of resource allocation by Mesos Master, individual frameworks can accept or reject offers based on the 2nd level scheduling policy and more specific resource constraints. For example, a “Bin Packing” algorithm will utilize more offers than a “First Fit” and “One Task per Cycle” task allocation policy.

So in a cycle, if Mesos Master allocates all the resources to a Framework-A, which uses “One Task Per Cycle” policy, then the majority of the allocated offers will be rejected 
and may be allocated to Framework-B, which uses the Bin-Packing policy to map tasks to offers.
To counter this resource distribution behavior, Framework-A needs more cycles to allocate more tasks than Framework-B. User level task allocation is one of the key attributes that drives a cluster towards unfair resource allocation. Other factors are (1) Refuse offer cycle,  (2) Resource holding period, (3) Task completion rate, and (4) Task arrival rate. We discuss how these attributes contribute towards the fairness of the overall resource distribution.

\begin{enumerate} 
\item \textbf{Refuse Offer Seconds:} This attribute defines the number of seconds for which resources from an agent cannot be offered again after the resources from the same agent were rejected or partially used. 

\item \textbf{Offer Holding Period:} A few frameworks hold resources for a specified period to make better scheduling decisions. An offer can be terminated from a framework when either a task is launched with the offer or the offer holding period is over. Mesos Master can also set the offer timeout and reclaim after the timer expires
If no timeout period is set, then the framework can hold an offer as long as it wants. A longer offer holding time decreases the resource utilization and can make other frameworks starve.

\item \textbf{Task Duration:} Long-running tasks block resources for a longer period and can make other users starve for resources. Apache Mesos does not kill tasks once they are started; it waits until the tasks finish even though other users may have more tasks to run.
  
\item \textbf{Task Arrival Rate:} Frameworks keep on launching tasks as long as the Mesos Master offers resources to them and there are tasks waiting in the queue. So, it is beneficial to frameworks that initiate tasks at a faster rate from a queue of tasks. A framework that launches tasks at a slower speed may face starvation because the resources are being blocked by the frameworks that are launching tasks at a higher rate.   
\end{enumerate}

\section{Evaluation and Experimental Results}
We ran experiments to determine the effect of various framework configurations and attributes on resource distribution and fairness. 
We used two off-the-shelf popular Mesos frameworks, Apache Aurora and Marathon, along with our MPI framework Scylla \cite{Saha2017Scylla:Jobs} to study and analyze the experimental results. Our experimental cluster has four nodes with a pool of 32 CPUs and 64 GBs of RAM. All tasks sent to the frameworks are identical in terms of resource requirements but in some experiments they differ in runtime. 

\begin{table}[h!]
\centering
\begin{tabular}{|l|p{0.60\linewidth}|} 
 \hline
 {\bf Software} & {\bf Version}\\
 \hline
 Ubuntu 		& Ubuntu 16.04.2 LTS (Xenial)\\ 
 \hline
 Apache Aurora 	& 0.17.0\\
 \hline
 Marathon & 1.4.0\\
 \hline
 Apache Mesos 	& 1.3.0\\
 \hline
 
\end{tabular}
\smallskip
\label{Software Stack and Vession}
\caption{\textit{Software Stack and Version}}
\vspace{-1.5em}
\end{table} 

\subsection{DRF based fairness on a multi-user cluster} \label{DRF based fairness on a multi-user cluster}
In this experiment, we ran two instances of Scylla (Scylla-A and Scylla-B), each with a queue of 100 tasks. Each task required $\langle 1 CPU, 1GB\ memory \rangle$ and runtime of 100 seconds. We first launched Scylla-A and waited for all its tasks to be launched on the cluster before launching Scylla-B. 

We can observe that after some fluctuations, the resource distribution is fair and both the frameworks are using close to 1/n of the cluster resources (which is 50\% in this case). In the cluster, for the requested configurations by the frameworks, at most 32 tasks can be launched. The fair share for each framework is 32 tasks. Figure~\ref{Scylla_NoRefuseSeconds} shows that each framework is running 10 to 20 tasks. This setup achieves fair distribution compared to the results shown in Figures~\ref{ScyllaVsMarathon} and \ref{ScyllaVsAurora}. 

\begin{figure}[h!]
 \vspace{-1.5em}
  \includegraphics[width=0.5\textwidth,height=4.5cm]
  {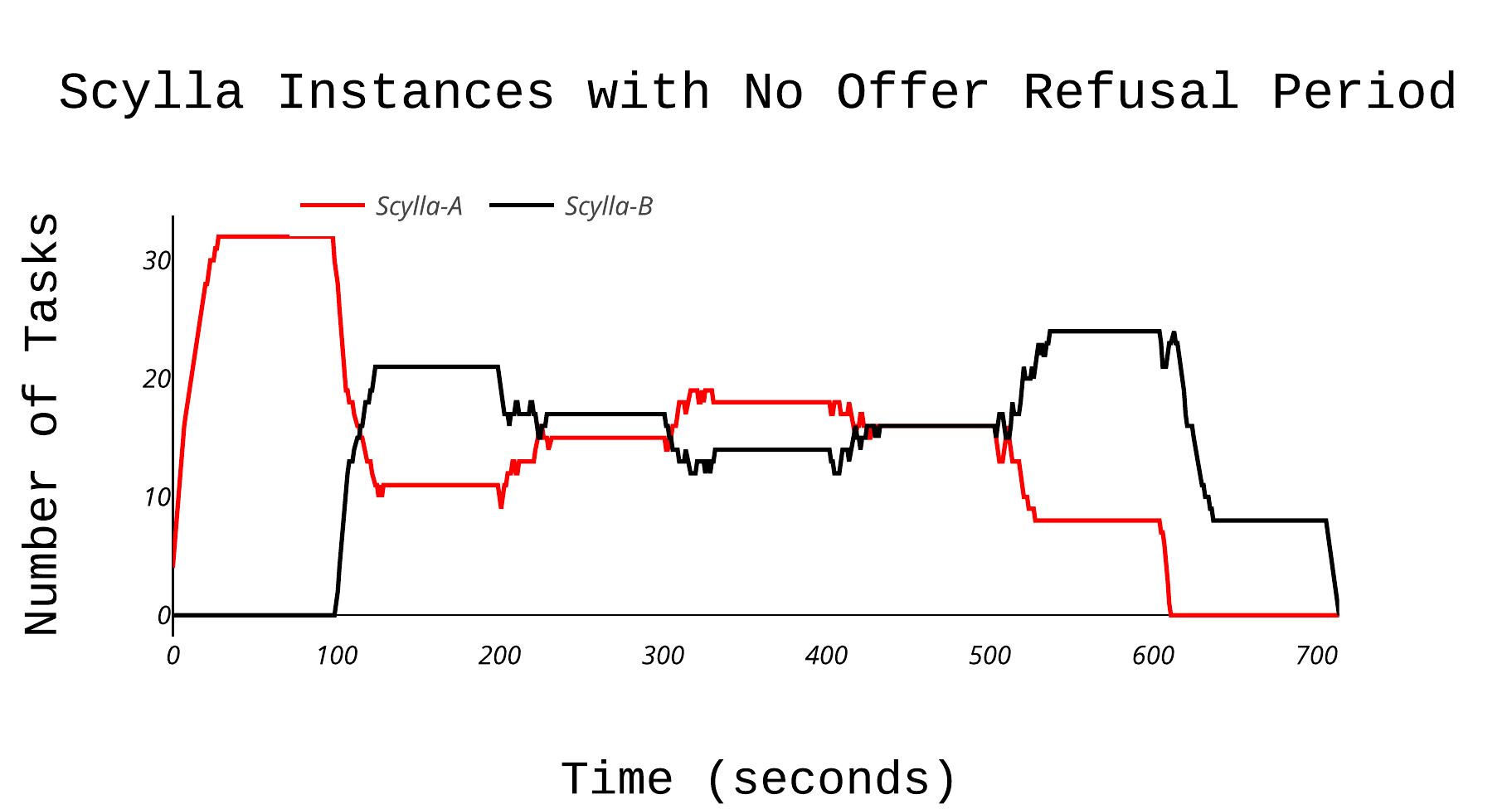}
  \caption{{\it Moderate Distribution: Number of tasks running every second by each Scylla instances setup with similar configurations. This setup has a moderate resource distribution among the frameworks when offer refusal period for both the Scylla instances is set to none.}
  }
  \label{Scylla_NoRefuseSeconds}
\end{figure}

\begin{figure}[h!]
 \vspace{-1.5em}
  \includegraphics[width=0.5\textwidth,height=4.5cm]
  {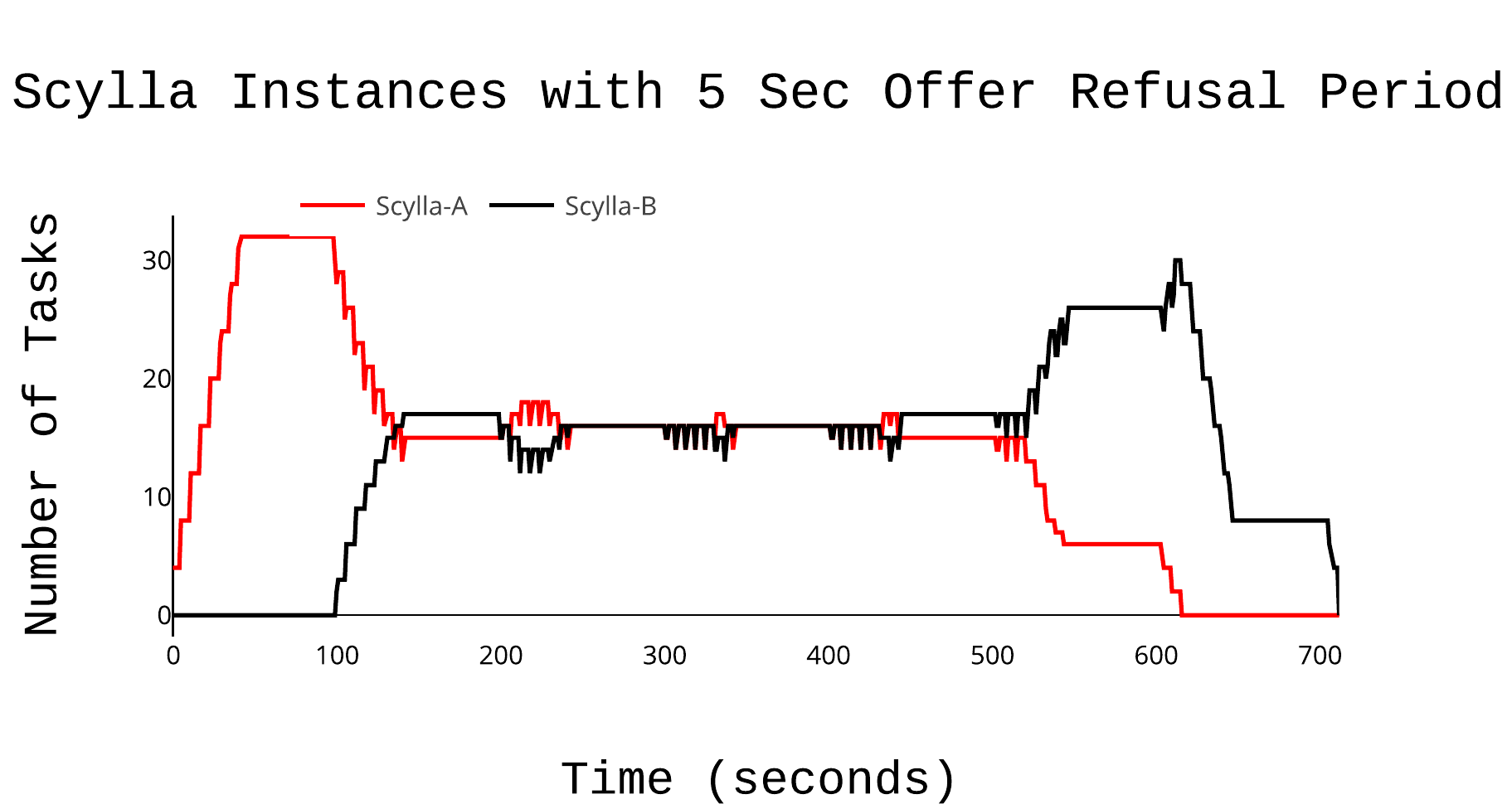}
  \caption{{\it Preferred Distribution: Number of tasks running every second by each Scylla instances setup with similar configurations. The refuse offer period is set to 5 seconds. 
  }}
  \vspace{-0.5em}
  \label{Scylla_5secRefuseSeconds}
\end{figure}

The Refuse Offer seconds attribute of one framework increases the opportunity for other frameworks to use resources that were rejected or partially used by a framework. So, to achieve better allocation, we ran the same experiment as Figure~\ref{Scylla_NoRefuseSeconds} but changed the Refusal Offer seconds by gradually increasing it from zero to five seconds. In Figure~\ref{Scylla_5secRefuseSeconds}, we can observe that both the frameworks are executing around 15-16 tasks for longer period time, which again maintains a better resource distribution compared to the previous experimental setup shown in Figure \ref{Scylla_NoRefuseSeconds}. We continued the experiment to  further increase the offer refusal seconds to 7 and 10 seconds, but did not see much improvement. So, in our experimental cluster, we kept the configuration of 5 seconds as the offer refusal period is optimal.


\subsection{Fairness with Marathon and Scylla Frameworks}
The second level scheduling policy can impact the clusters resource distribution and lead to unfair distribution of resources. For this experiment, we deployed Marathon and Scylla, wherein Scylla employs First-Fit scheduling policy. Both the frameworks were given a queue with 100 tasks, each requiring 1 CPU and 1 GB of RAM.

\begin{figure}[h!]
 \vspace{-1em}
  \includegraphics[width=0.5\textwidth,height=4.5cm]
  {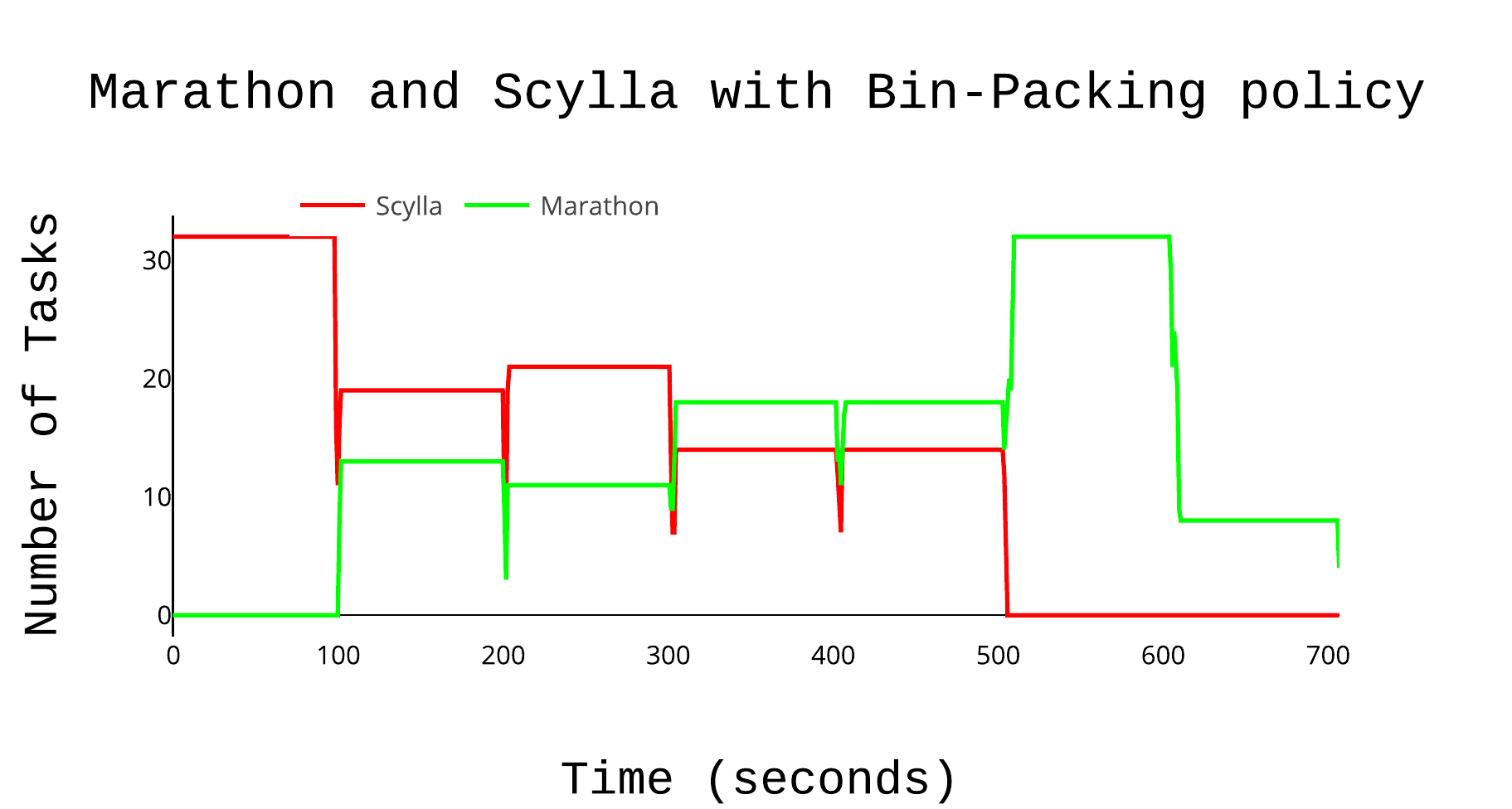}
  \caption{{\it Moderate Distribution: Marathon and Scylla are vying for resources in a Mesos cluster when Scylla is configured with Bin-Packing as the second level task allocation policy. This resource distribution is moderately better compared to the unfair distribution we noticed in Figure \ref{ScyllaVsMarathon}.
  }}
  \label{ScyllaBPR3VsMarathon}
\end{figure}

\begin{figure}[h!]
 \vspace{-1em}
  \includegraphics[width=0.5\textwidth,height=4.5cm]
  {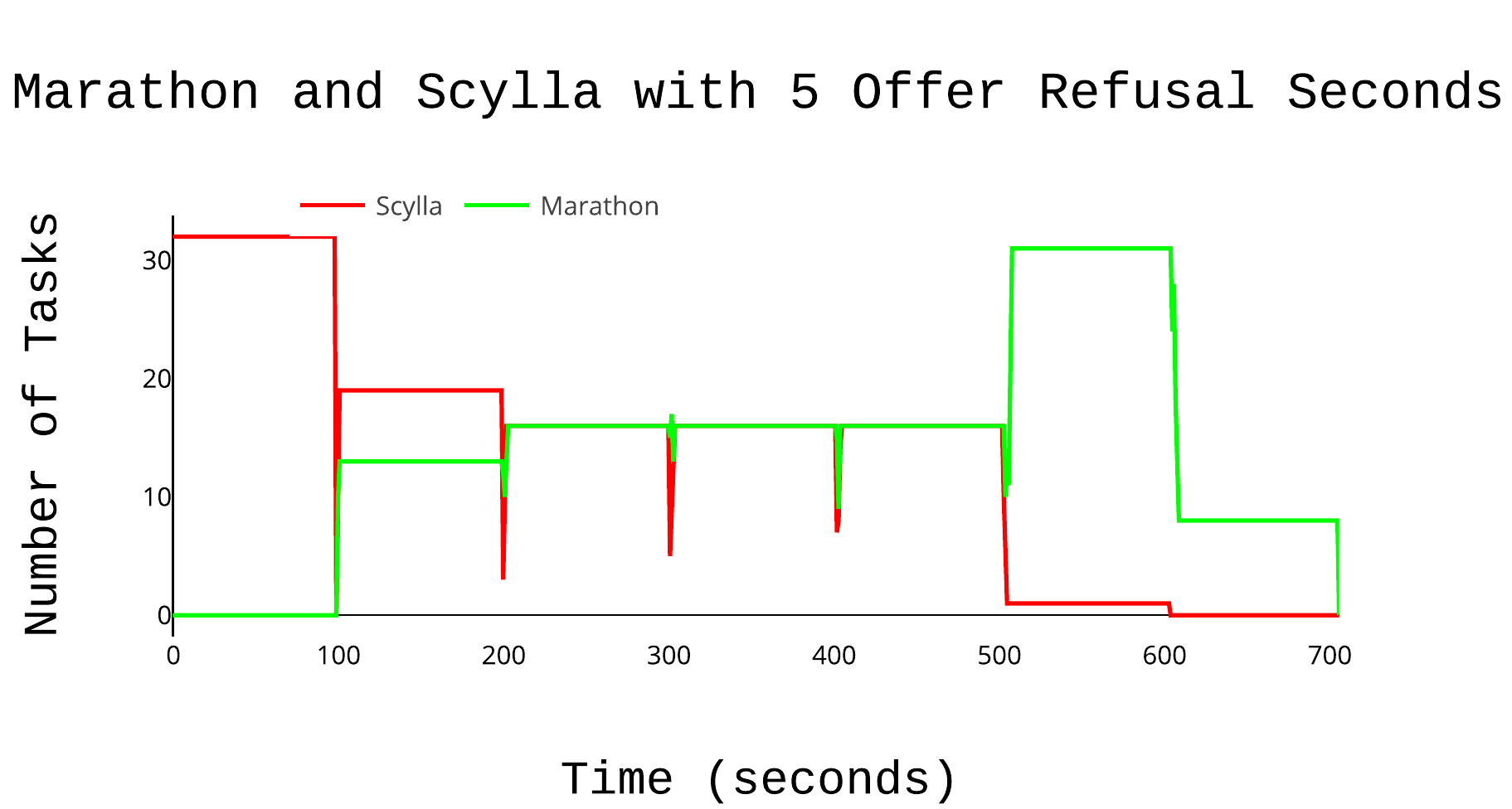}
  \caption{{\it Preferred Distribution: Marathon and Scylla are vying for resources in a Mesos cluster when Scylla is configured with Bin-Packing as second level task allocation policy and 5 second refuse offer period is set as a framework attribute of Scylla, for better resource distribution.
  }}
  \vspace{-0.5em}
  \label{ScyllaBPR5VsMarathon}
\end{figure}

In Figure \ref{ScyllaVsMarathon}, we can observe how Marathon's greedy resource consumption policy consumes more resources and launches more tasks even though another framework was waiting for resources. Marathon used resources that exceeded the fair share, whereas Scylla received a smaller share. This greedy approach of Marathon caused a 38\% reduction in fair resource allocation to Scylla, as shown in Figure \ref{ScyllaVsMarathon}. We changed the configuration and instantiated Scylla with Bin-Packing as second level scheduling policy. In Figure \ref{ScyllaBPR3VsMarathon}, we can see how each framework started using close to fair share of the cluster resources. Due to Scylla's Bin-Packing policy, it receives 5\% more resources than the fair share limit, which is better than the loss of 38\% fairness in allocation shown in Figure~\ref{ScyllaVsMarathon}. This situation is further improved to 3\% above the fair share limit by increasing the offer refusal period to 5 seconds, which was previously determined to be the optimal value for this cluster in Figure \ref{ScyllaBPR5VsMarathon}.


\subsection{Fairness on Apache Aurora and Scylla based cluster}

Figures~\ref{auroraAndScyllaR0}, \ref{auroraAndScyllaR5} and \ref{auroraAndScyllaR20} show how we can gradually achieve better resource distribution with Apache Aurora and Scylla. With Apache Aurora, we faced some challenges in attaining fairness as it holds all the offered resources for a specified period. When a framework holds resources for a while, even if it does not launch any tasks, those held offers are weighed against the framework in computing the dominant share of the framework. Due to the increment of dominant share, Aurora's priority to receive further offers goes down, and another framework gets those offers. 

To explain this scenario in more detail, consider the following example -- a cluster with four nodes, each one containing the following resource configuration: 8 CPUs, 16GB of memory, and 32GB of disk space. Mesos Master will receive advertisement of offers from all the Agents in the following manner:
\begin{itemize} 
\item Offer 1 $\langle 8 CPU, 16GB\ memory, 32000MB\ disk\rangle$
\item Offer 2 $\langle 8 CPU, 16GB\ memory, 32000MB\ disk\rangle$
\item Offer 3 $\langle 8 CPU, 16GB\ memory, 32000MB\ disk\rangle$
\item Offer 4 $\langle 8 CPU, 16GB\ memory, 32000MB\ disk\rangle$
\end{itemize} 

Now consider Framework A launches 32 tasks, each $\langle 1CPU, 2GB\ memory, 100MB\ disk\rangle$, in the cluster before Framework B is started. Table \ref{my-label1} shows the current status of the cluster where we can see that it is 100\% occupied by framework A, for CPU and memory resources, and exhibits a dominant share of 100\%. As a result, any available resources should subsequently go to framework B. During this period there will be resource offers like $\langle 0CPU, 0GB\ memory, 31200MB\ disk \rangle$. Let us consider that we have framework B, which holds resources for a specified amount of time hoping to make better task allocation in its internal scheduling. During this offer holding period, let us say framework A completes 4 tasks and as a result a total offer of $\langle 4cpu,8gb-mem, 400gb-disk \rangle$ will be freed. Mesos Master will then allocate the offers through next DRF cycle. Mesos Master will compute the resource share of each framework as shown in Table \ref{my-label2}.

\begin{table}
	\centering
	\caption{\it {100\% CPU and Memory resource is consumed}} 
	\label{my-label1}
	\begin{tabular}{|c|c|c|}
    	\hline
			CPU   & Memory & Disk    \\
        \hline
			100\% & 100\%  & 2.5\%   \\
        \hline
	\end{tabular}
\end{table}

\begin{table}
    \centering
    \caption{\it{Resource allocation in presence of both frameworks}}
	\label{my-label2}
	\begin{tabular}{|c|c|c|c|} 
    	\hline
			Framework & CPU & Memory & Disk  \\ [0.5ex] 
 		\hline
			Framework-A & 87\% & 87\% & 2.8\%   \\
		\hline
			Framework-B & 0\% & 0\% & 97.81\%  \\
 		\hline
	\end{tabular}
\end{table}


\begin{table}
	\centering
	\caption{\it{100\% Disk Resource is Allocated}}
	\label{my-label3}
	\begin{tabular}{|c|c|c|}
    	\hline
 			CPU & Memory & Disk   \\
 		\hline
            100\% & 100\% & 100\%   \\
		\hline
    \end{tabular}
\end{table}

Now any framework that has a lower dominant share will get the opportunity to use these available offers before framework A acquires it. Framework B’s CPU and memory share is 0\%, and disk share is 97.81\%. Even though Framework B is not launching any tasks, due to the disk resource on hold, Mesos Master will determine its dominant (disk) share to be 97.81\%, which is higher than frameworks A's dominant share. So framework A gets the chance to use the available offers before framework B. This causes starvation for framework B until framework A is done with executing all the tasks. 

To demonstrate with an experiment, we setup a cluster with the same resource configuration and used Scylla as Framework A. We deployed Apache Aurora as framework B, which holds resources for 5 minutes by default~\cite{AuroraConfiguration}. 
In Figure~\ref{auroraAndScyllaR0} we can see that Aurora is able to launch less number of tasks in the presence of Scylla as Aurora faces 89\% fairness reduction in resource allocation. Scylla receives more CPU and memory resources and is able to launch all the tasks ahead of Aurora.
\begin{figure}[h!]
 \vspace{-1em}
  \includegraphics[width=0.5\textwidth,height=4.5cm]
  {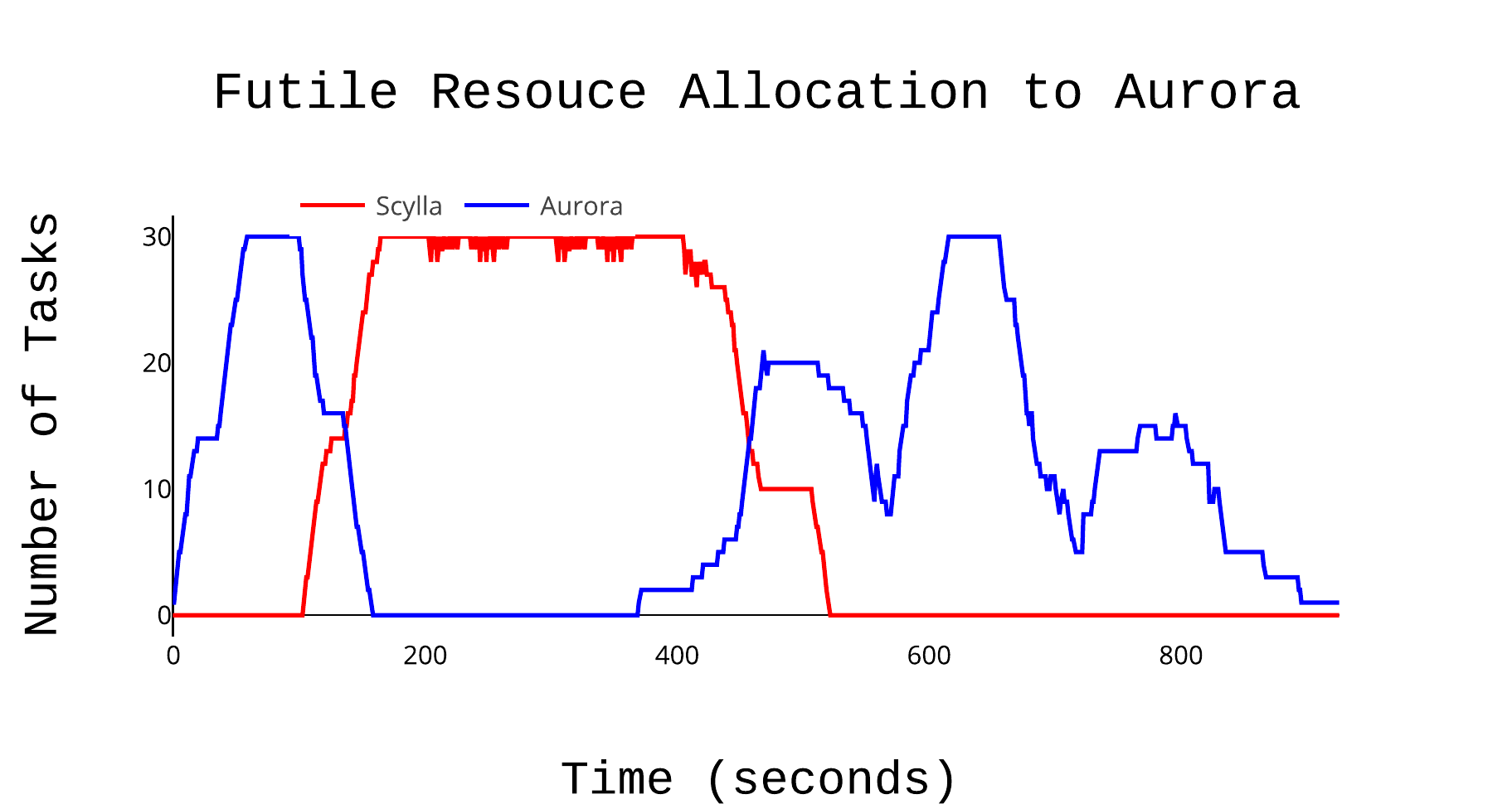}
  \caption{{\it Poor Distribution: Apache Aurora is launching a small number of tasks in the presence of Scylla. Entire cluster's resources are being used by Scylla to launch most of its tasks, which is leading to poor resource distribution.}}
  \label{auroraAndScyllaR0}
\end{figure}

\begin{figure}[h!]
 \vspace{-1.5em}
  \includegraphics[width=0.5\textwidth,height=4.5cm]
  {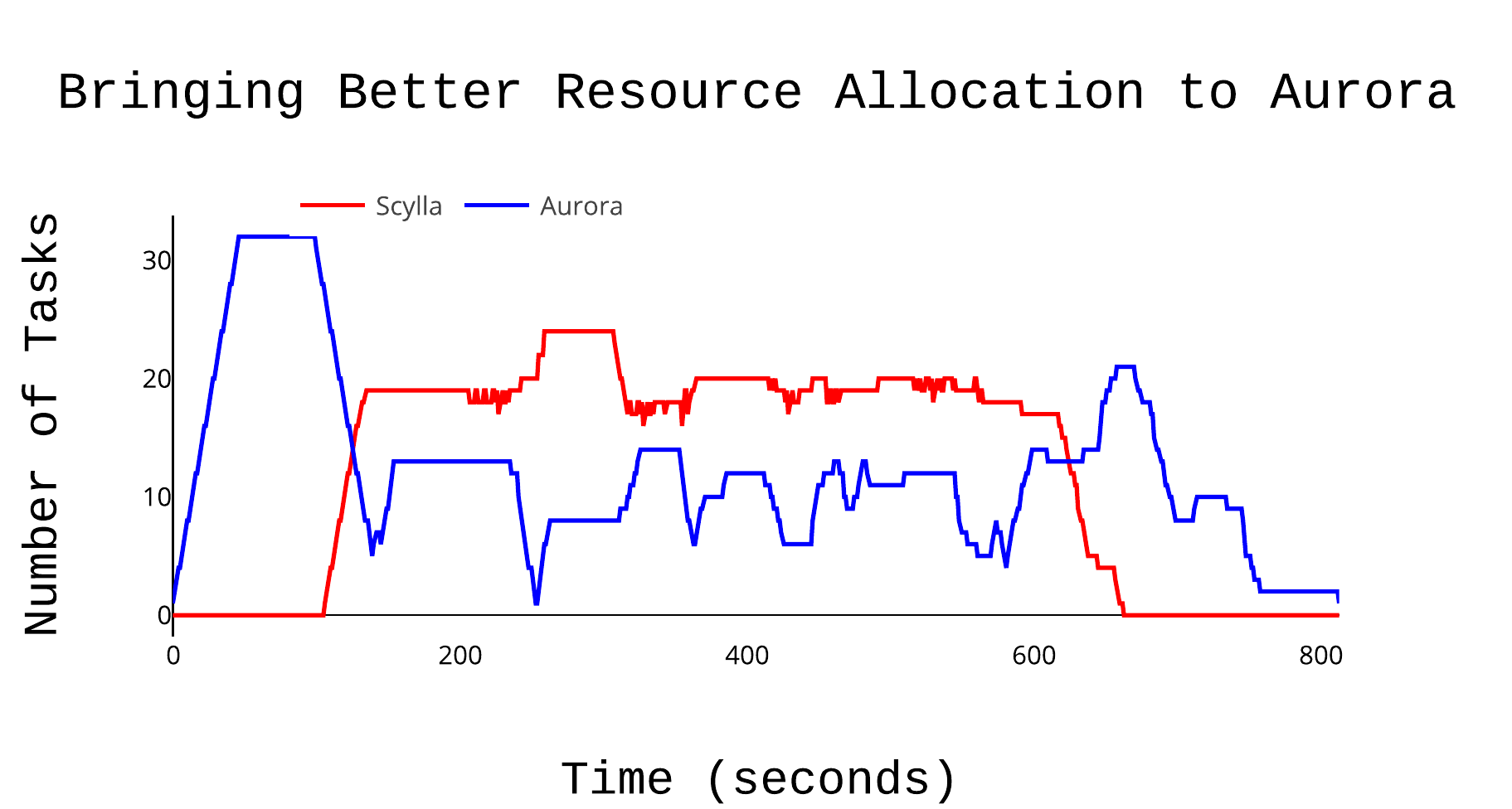}
  \caption{{\it Better Distribution: Resource distribution is improved between Apache Aurora and Scylla by addressing the problem due to Aurora's resource holding feature.}}
  \label{auroraAndScyllaR5}  
\end{figure}

\begin{figure}[h!]
 \vspace{-1.5em}
  \includegraphics[width=0.5\textwidth,height=4.5cm]
  {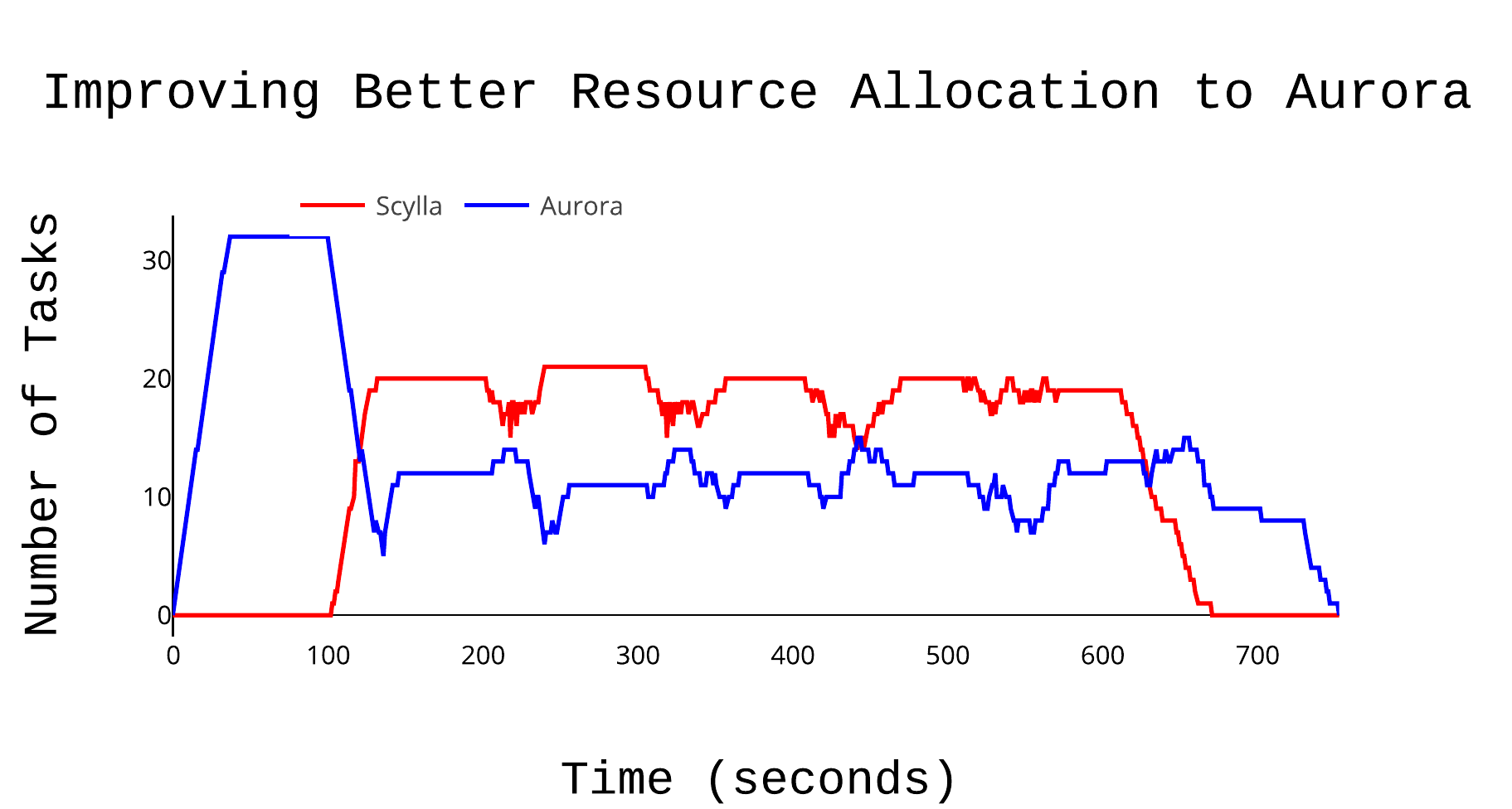}
  \caption{{\it Improved Distribution: Increased offer refusal period of Scylla improves resource distribution to Aurora.}}
  \label{auroraAndScyllaR20}
  \vspace{-1.5em}
\end{figure}

To improve resource distribution, we tried to reduce the free disk resource in the cluster. Instead of launching tasks with  $\langle 1CPU, 2GB\ memory, 100MB\ disk\rangle$, we launched each task with $\langle 1CPU, 2GB\ memory, 4096MB\ disk\rangle$ to combine more disk resources in the task requirements. Now, at the beginning when Scylla is running 32 tasks, its resource share metric will be as shown in Table~\ref{my-label3}. As there exist no offers due to the unavailability of resources, Aurora will not receive offers with big disk space like in the previous case. As Aurora is not holding any resources, any freed up resources will be offered to it, which will bring a better resource distribution in the cluster. 

Figure~\ref{auroraAndScyllaR5} shows a comparatively better resource distribution where Aurora has 35\% reduction in fairness. We have seen in previous experiments (see Figures \ref{Scylla_5secRefuseSeconds} and \ref{ScyllaBPR5VsMarathon}) an increment of Refuse Offer seconds of a framework improves the chances of another framework to get relatively better resources. Figure \ref{auroraAndScyllaR20} shows relatively better results than Figure \ref{auroraAndScyllaR5} and unfairness is reduced to 28\% for Aurora.

\subsection{Impact of Idle Users on Resource Distribution}
In this experiment, we study how the presence of idle frameworks in a Mesos cluster can cause low resource utilization. We launched a single instance of Scylla and recorded the time it takes to complete 100 tasks. We launched all the tasks in a fixed interval of two seconds throughout these experiments. Next, we increased the number of framework instances and observed how the makespan increases, in presence of other idle frameworks, due to the DRF resource allocation policy. 

\begin{figure}[h!]
 \vspace{-1.5em}
  \includegraphics[width=0.5\textwidth]
  {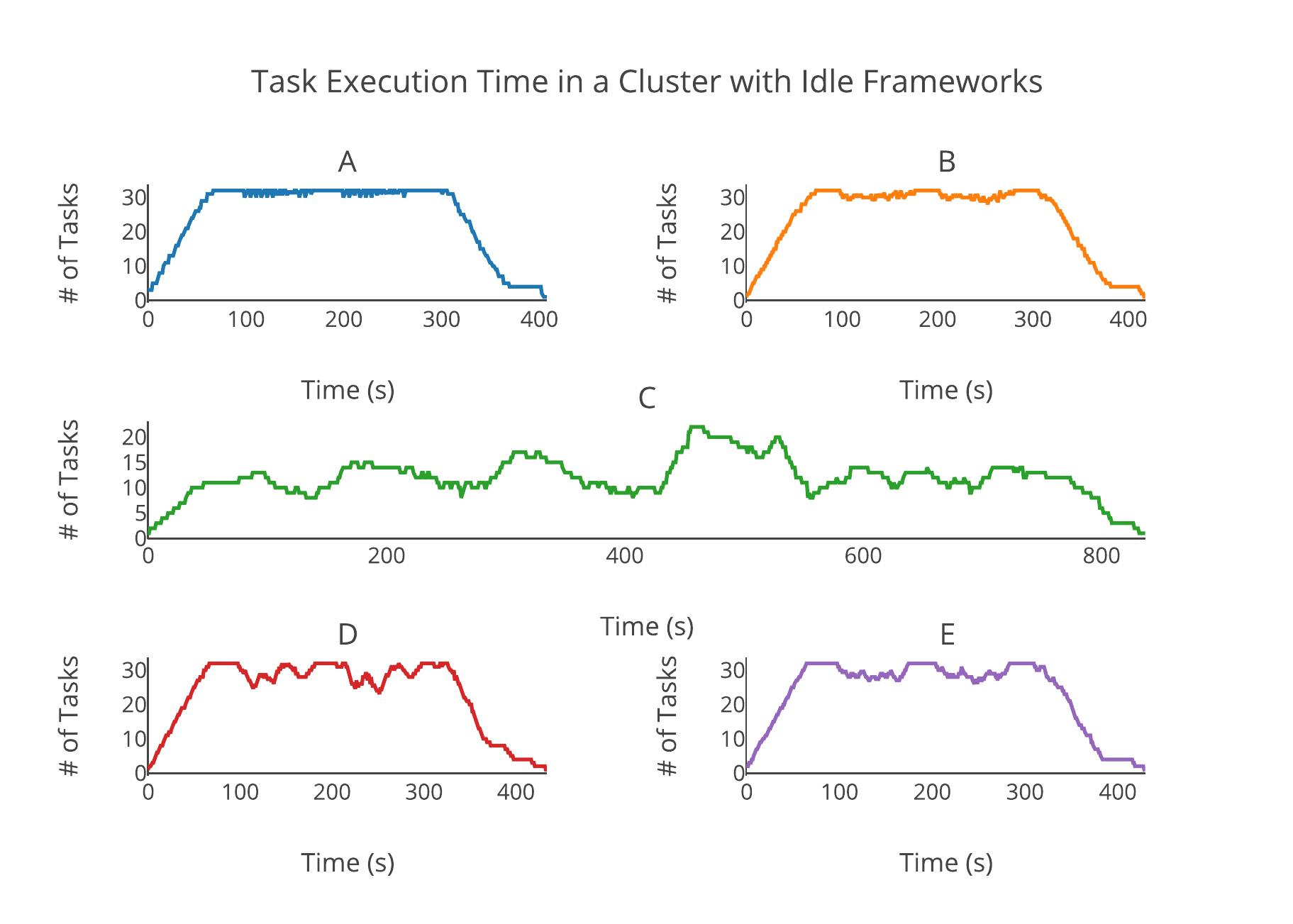}
  \caption{{\it Impact of Idle Frameworks: The required time to launch all the tasks of an active framework rises as the number of idle frameworks in the cluster increase. 
This could be addressed by increasing the offer refusal period of idle frameworks.
}}
\vspace{-1em}
  \label{idleGraphs}
\end{figure}

Figure~\ref{idleGraphs} shows how makespan increases as we increase the number of idle frameworks. From Figure~\ref{idleGraphs}A to Figure~\ref{idleGraphs}C, we increased the number of idle frameworks in the cluster and observed that the makespan doubles when the number of idle framework goes up to five.

DRF allocation favors the framework that has received less resource allocations. Thus, Mesos Master offers the resources to the idle framework in the current cluster environment even though these frameworks do not have any tasks to launch. Their dominant share is 0\%. DRF fairness always prefers framework with the lower dominant share and this can cause starvation for a framework that has a pending list of tasks to launch. To avoid starvation of an active framework, we can increase the offer refusal duration for the idle frameworks that do not have pending tasks. When a task appears on those idle framework, the filter can be removed for it to accept offers for the pending tasks. To study this case, we increased the offer refuse duration of the idle frameworks from five to 10 seconds, and reduced the active frameworks' offer refuse duration to two seconds. We can observe in Figures~\ref{idleGraphs}D and \ref{idleGraphs}E how this change improves the makespan. Any further reduction in the offer refusal period for the active framework, however, did not produce further improvements. 


\subsection{Long running tasks towards unfair distribution}
Mesos lets a framework exceed its fair share if the lower share framework does not want the resources. As Mesos does not revoke resources until a task is completed, a framework may have to wait for those resources to be released before it can get its share. Thus, improvement in utilization comes at the cost of not providing a guarantee that a user can get its fair share without waiting. Mesos allows the use of a {\tt quota}, for role-based static and dynamic reservation of resources, to ensure that a framework will always get its share no matter what is being requested by other frameworks. However, the use of quota can lead to underutilization of the cluster resources if a reserved resource is not being used by a framework. 

\begin{figure}[h!]
 \vspace{-1em}
  \includegraphics[width=0.5\textwidth,height=4.5cm]
  {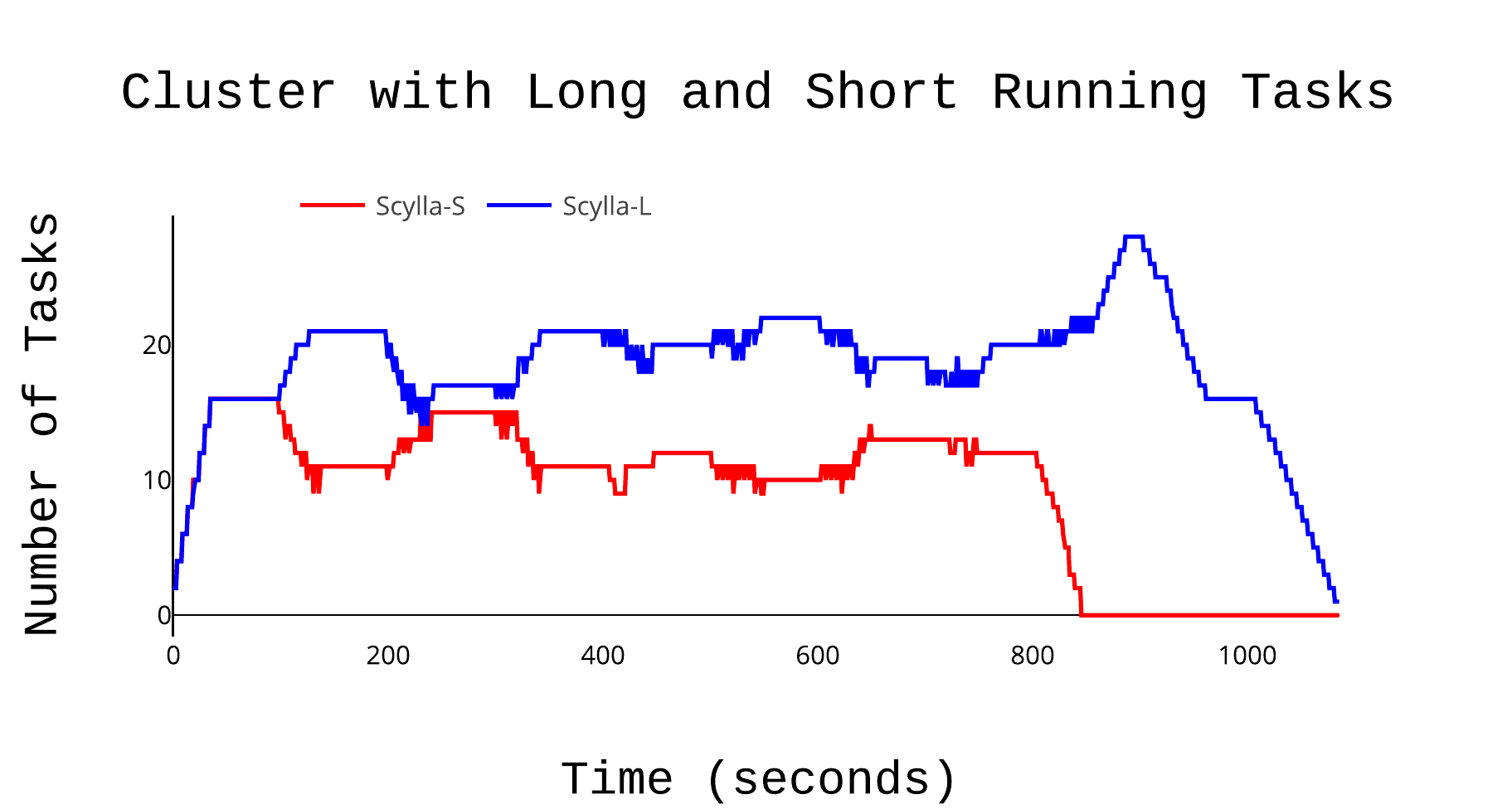}
  \caption{{\it  Poor Distribution: Resource distribution when two identical frameworks launch tasks of different durations but at the same launching rate.}}
  \label{delay_longDelay5_shortDelay5}
  \vspace{-1.5em}
\end{figure}

\begin{figure}[h!]
\vspace{-0.5em}
  \includegraphics[width=0.5\textwidth,height=4.5cm]
  {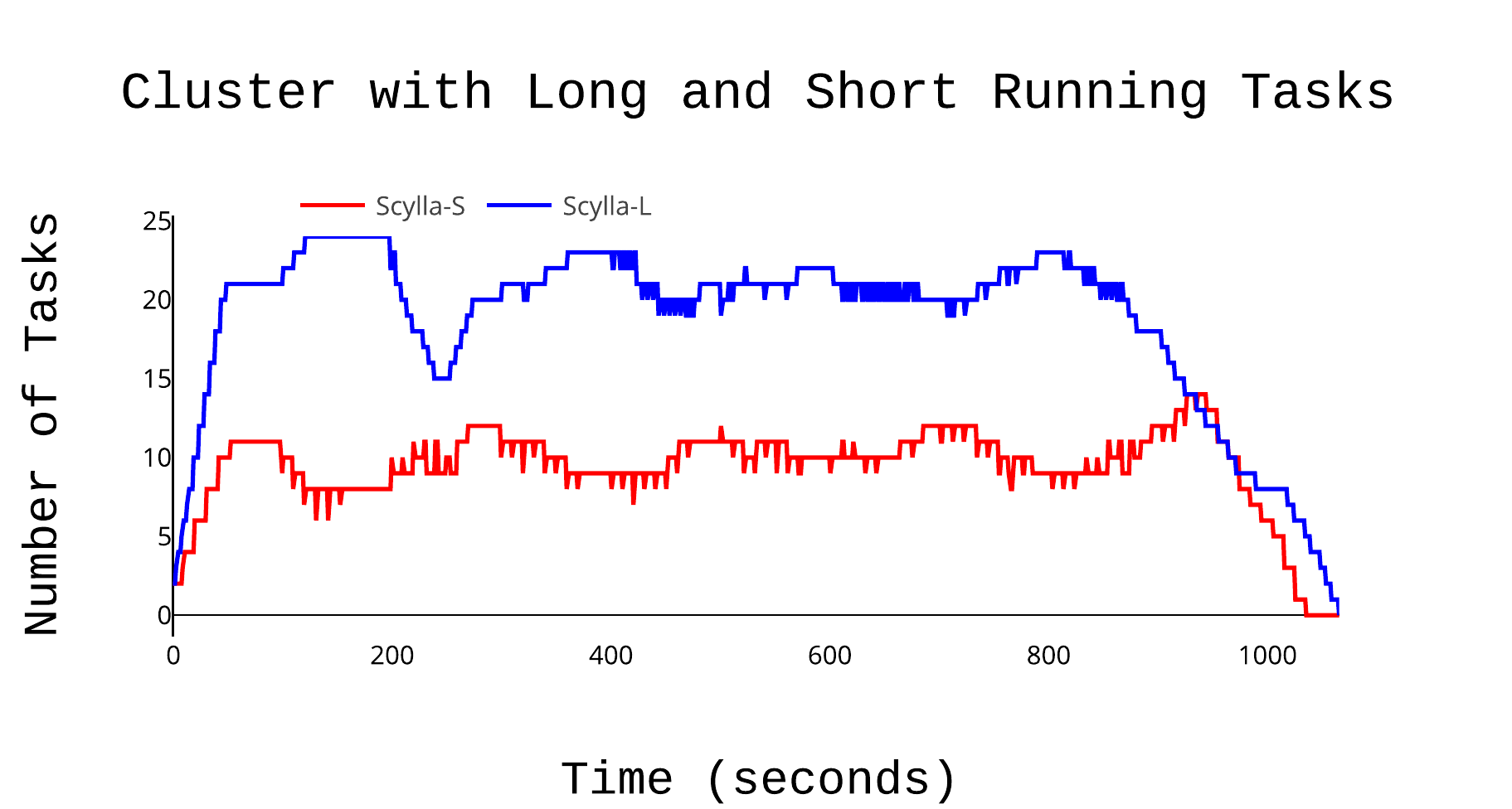}
  \caption{{\it Worse Distribution: Cluster with long and short Resource distribution scenario when long running tasks are launched at a higher rate and short running tasks are launched at a slower rate.}}
  \label{delay_longDelay5_shortDelay10}
    \vspace{-0.5em}
\end{figure}

\begin{figure}[h!]
 \vspace{-1em}
  \includegraphics[width=0.5\textwidth,height=4.5cm]
  {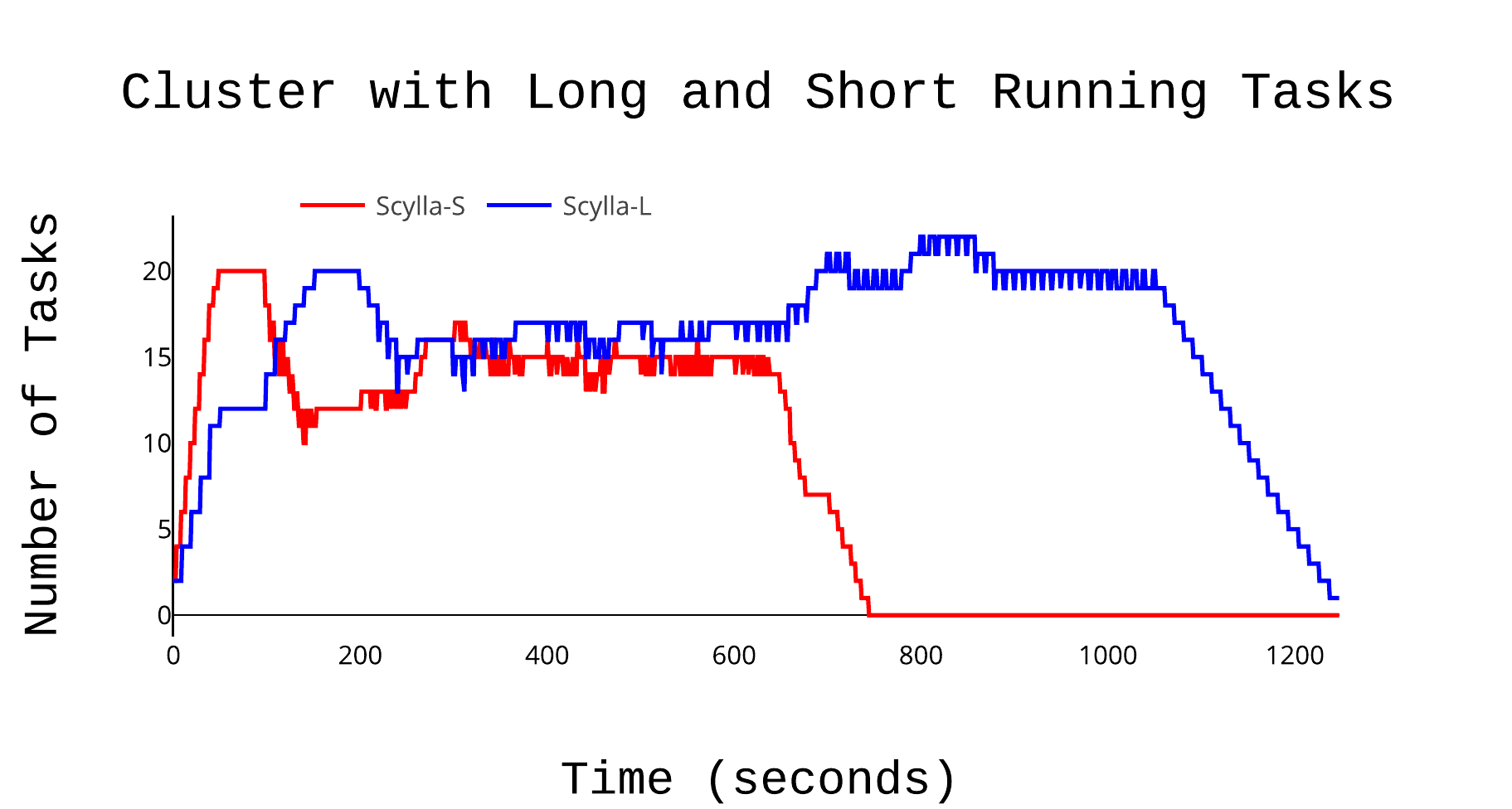}
  \caption{{\it Improved Distribution: Cluster with long and short running tasks. This figure illustrates an improved resource distribution when long running tasks are launched at a slow rate and short running tasks are launched at a high rate.}}
  \label{delay_longDelay10_shortDelay5} 
  \vspace{-1.5em}
\end{figure}

In this experiment, we setup two instances of the Scylla framework with similar configuration (Scylla-L and Scylla-S). Scylla-L launches long-running tasks whereas Scylla-S runs short-running tasks. Both the frameworks allocate tasks with “First-Fit” as the second level scheduling policy and receive tasks in an interval of 5 seconds. Short running tasks required 1CPU and 1GB of memory resources and run for 100 seconds. Whereas, long running tasks required similar amount of resources but run for 200 seconds. We launched a total of 100 short running tasks and only 50 long running tasks. As the runtime of long running tasks is double that of the short running tasks, it keeps the total runtime of both the types of tasks the same. Our experimental results in Figure~\ref{delay_longDelay5_shortDelay5} show how this experiment results in an unfair distribution in the cluster. Framework-S starves for a longer period in the Mesos cluster as it receives 23\% unfair resource allocation.

To observe how this resource distribution is impacted by task arrival rate, we changed the frequency at which tasks appear in the queue for each framework. Figure~\ref{delay_longDelay5_shortDelay10} shows how resources are distributed among frameworks when Framework-L launches long-running tasks every 5 seconds and Framework-S launches tasks every 10 seconds. The lower task arrival rate of Framework-S results in 37\% reduction in fairness. To improve the scenario, we interchanged the task arrival rate of both the frameworks. We see an improvement in the resource distribution shown in Figure~\ref{delay_longDelay10_shortDelay5} where the unfairness is reduced to 7\% for Framework-S. 


\section{Related Work}
The work on evaluation of DRF is in its early stages.

Ghodi et al. \cite{ghodsi2011dominant}  introduced DRF as a generalization of the well-known Max-Min traditional problem.
Our work complements their research as we are evaluating how the implementation of DRF in Apache Mesos differs from the proposed DRF, and how framework policies affect resource distribution.

Dimopoulos et al.~\cite{DimoPerformanceClouds} show how big data frameworks (Hadoop, Spark, and Storm) hinder each other in a Mesos cluster under resource constraints. They compare the frameworks with different data sizes to see how performance varies with data volume.

Wang et al. \cite{wang2014dominant} generalized DRF to work on multiple heterogeneous servers.
Saha et al.~\cite{Saha2016IntegratingMesos} show how Apache Mesos can be integrated into a scientific cluster to leverage DRF based fair resource distribution. A Mesos framework, Scylla \cite{Saha2017Scylla:Jobs}, was developed to orchestrate scientific MPI tasks on a Mesos cluster. 
\section{Conclusion}


\begin{itemize}
\item 
The DRF based allocation module may not provide enough resources to frameworks such as Apache Aurora, which holds on to resources instead of immediately using them.
Framework specific attributes, like offer holding period, are critical and informed decisions based on the existence of other frameworks can reduce an unfair allocation from 90\% to 28\%.


\item{Frameworks can refuse to accept offers from the Mesos allocation module for a configurable period of time. This offer refusal period increases the opportunity for other frameworks to use the refused offer, and in turn leads to better resource distribution. For a change in offer refusal period from 0 to 5 seconds, the gains in fairness range can from 85\% to 99\%.}

\item{The second level scheduling policy of a framework can incorporate greediness and make other frameworks starve for resources. In such cases, competing frameworks that allow change in their scheduling policy, to bin packing for example, can reduce the unfairness from 38\% to just 3\%.
}

\item{Once Mesos launches tasks, it does not terminate them even if the frameworks' dominant share exceeds the fair share limit of the cluster. This feature can lead other frameworks to starve.
Increased arrival rate of short running tasks and decreased arrival rate of long-running tasks from two different frameworks respectively can reduce the unfair resource distribution from 23\% to 7\%. }

\end{itemize}

\bibliographystyle{IEEEtran}
\bibliography{Mendeley,bibbandwidth}

\end{document}